\documentclass[iop]{emulateapj}
\usepackage{apjfonts}
\usepackage{graphicx}
\usepackage{color}

\def\d3{$\delta_{3}$ }
\def\1d3{$(1 + \delta_{3})$ }
\def\l1d3{$\log_{10}(1 + \delta_{3})$ }

\def\s3{$\Sigma_{3}$}
\def\ha{H$\alpha$}
\def\hb{H$\beta$}
\def\oone{[OI] 6300}

\def\othree{[OIII] 5007}
\def\ntwo{[NII] 6584}
\def\stwo{[SII] 6717,6731}

\def\24m{24 $\mu$m}

\def\kms{${\rm km~s^{-1}}$ }

\def\Msolar{$\rm M_{\odot}$}

\def\rmxaa{RMxAA}
\def\ser{{S\'{e}rsic\ }}

\shorttitle{A MaNGA discovered \ha~blob}
\shortauthors{Lin et al.}

\begin{document}

\title{SDSS IV MaNGA: Discovery of an \ha~blob associated with a dry galaxy pair --  ejected gas or a `dark' galaxy candidate?}

\author{Lihwai Lin \altaffilmark{1}, Jing-Hua Lin \altaffilmark{1,2}, Chin-Hao Hsu \altaffilmark{1,2}, Hai Fu \altaffilmark{3}, Song Huang \altaffilmark{4}, Sebasti\'{a}n F. S\'{a}nchez \altaffilmark{5}, Stephen Gwyn\altaffilmark{6}, Joseph D. Gelfand \altaffilmark{7,8}, Edmond Cheung \altaffilmark{4}, Karen Masters \altaffilmark{9}, S\'{e}bastien Peirani \altaffilmark{10,4}, Wiphu Rujopakarn\altaffilmark{4,11}, David V. Stark \altaffilmark{4}, Francesco Belfiore \altaffilmark{12,13}, M. S. Bothwell \altaffilmark{12,13}, Kevin Bundy \altaffilmark{14,4}, Alex Hagen \altaffilmark{15,16}, Lei Hao \altaffilmark{17}, Shan Huang \altaffilmark{8}, David Law \altaffilmark{18}, Cheng Li \altaffilmark{19}, Chris Lintott \altaffilmark{20}, Roberto Maiolino \altaffilmark{12,13}, Alexandre Roman-Lopes \altaffilmark{21}, Wei-Hao Wang \altaffilmark{1}, Ting Xiao \altaffilmark{17}, Fangting Yuan \altaffilmark{17}, Dmitry Bizyaev \altaffilmark{22,23}, Elena Malanushenko \altaffilmark{22}, Niv Drory \altaffilmark{24}, J. G. Fern\'{a}ndez-Trincado \altaffilmark{25}, Zach Pace \altaffilmark{26}, Kaike Pan \altaffilmark{22}, Daniel Thomas \altaffilmark{9}}

\altaffiltext{1}{Institute of Astronomy \& Astrophysics, Academia Sinica, Taipei 10617, Taiwan; Email: lihwailin@asiaa.sinica.edu.tw}%lihwai, jing-hua, chin-hao, wei-hao
\altaffiltext{2}{Department of Physics, National Taiwan University, 10617, Taipei, Taiwan}%jing-hua, chin-hao
\altaffiltext{3}{Department of Physics \& Astronomy, University of Iowa, Iowa City, IA 52242}%Hai
\altaffiltext{4}{Kavli Institute for the Physics and Mathematics of the Universe
(WPI), The University of Tokyo Institutes for Advanced Study, The
University of Tokyo, Kashiwa, Chiba 277-8583, Japan}%Edmond, Kevin, wiphu, david stark
\altaffiltext{5}{Instituto de Astronom\'{i}a, Universidad Nacional Auton\'{o}ma de Mexico, A.P. 70-264, 04510, M\'{e}xico, D.F., M\'{e}xico}% sanchez
\altaffiltext{6}{NRC-Herzberg Astronomy and Astrophysics, National Research Council of Canada, 5071 West Saanich Road, Victoria, British Columbia V9E 2E7, Canada}% gwyn
\altaffiltext{7}{NYU Abu Dhabi, P.O. Box 129188, Abu Dhabi, UAE}%yosi
\altaffiltext{8}{Center for Cosmology and Particle Physics, New York University, New York, NY 10003, USA}%Yosi, shan
\altaffiltext{9}{Institute of Cosmology \& Gravitation, University of Portsmouth, Dennis Sciama Building, Portsmouth, PO1 3FX, UK}% karen masters
\altaffiltext{10}{Institut d'Astrophysique de Paris (UMR 7095: CNRS \& UPMC), 98 bis Bd
Arago, 75014 Paris, France}%peirani
\altaffiltext{11}{Department of Physics, Faculty of Science, Chulalongkorn
University, 254 Phayathai Road, Pathumwan, Bangkok 10330, Thailand}
\altaffiltext{12}{Cavendish Laboratory, University of Cambridge, 19 J. J. Thomson Avenue, Cambridge CB3 0HE, United Kingdom}%Francesco, roberto, matt
\altaffiltext{13}{University of Cambridge, Kavli Institute for Cosmology, Cambridge, CB3 0HE, UK.}%Francesco, roberto, matt
\altaffiltext{14}{UCO/Lick Observatory, University of California, Santa Cruz, 1156 High St. Santa Cruz, CA 95064, USA}%Kevin
\altaffiltext{15}{Dept. of Astronomy \& Astrophysics, Pennsylvania State
University, University Park, PA 16802, USA}
\altaffiltext{16}{Institute for Gravitation and the Cosmos, Pennsylvania State University, University Park, PA 16802, USA}%Alex
\altaffiltext{17}{Shanghai Astronomical Observatory, Chinese Academy of Science, 80 Nandan Road, Shanghai 200030, China}%Lei, Ting, Xiaoting
\altaffiltext{18}{Space Telescope Science Institute, 3700 San Martin Drive, Baltimore, MD 21218, USA}%david law
\altaffiltext{19}{Tsinghua Center of Astrophysics \& Department of Physics, Tsinghua University, Beijing 100084, China}%cheng
\altaffiltext{20}{Sub-department of Astrophysics, Department of Physics, University of Oxford, Denys Wilkinson Building, Keble Road, Oxford OX1 3RH}%chris
\altaffiltext{21}{Departamento de Fisica, Facultad de Ciencias, Universidad de La Serena, Cisternas 1200, La Serena, Chile}%Alexandre 
\altaffiltext{22}{Apache Point Observatory and New Mexico State
University, P.O. Box 59, Sunspot, NM, 88349-0059, USA}
\altaffiltext{23}{Sternberg Astronomical Institute, Moscow State
University, Moscow, Russia}
\altaffiltext{24}{McDonald Observatory, The University of Texas at Austin, 1 University Station, Austin, TX 78712, USA}
\altaffiltext{25}{Institut Utinam, CNRS UMR 6213, Universit\'e de Franche-Comt\'e, OSU THETA Franche-Comt\'e-Bourgogne, Observatoire de Besan\c{c}on, \\BP 1615, 25010 Besan\c{c}on Cedex, France}
\altaffiltext{26}{Department of Astronomy, University of Wisconsin-Madison, 475N. Charter St., Madison WI 53703, USA}

\begin{abstract}
We report the discovery of a mysterious giant \ha~blob that is $\sim 8$ kpc away from the main MaNGA target 1-24145, one component of a dry galaxy merger, identified in the first-year SDSS-IV MaNGA data. The size of the \ha~blob is $\sim$ 3-4 kpc in radius, and the \ha~distribution is centrally concentrated. However, there is no optical continuum counterpart in deep broadband images reaching $\sim$26.9 mag arcsec$^{-2}$ in surface brightness.  
We estimate that the masses of ionized and cold gases are $3.3 \times 10^{5}$ \Msolar~and $< 1.3 \times 10^{9}$ \Msolar, respectively. 
The emission-line ratios indicate that the \ha~blob is photoionized by a combination of massive young stars and AGN. Furthermore,
the ionization line ratio decreases from MaNGA 1-24145 to the \ha~blob, suggesting that the primary ionizing source may come from MaNGA 1-24145, likely a low-activity AGN. Possible explanations of this \ha~blob include AGN outflow, the gas remnant being tidally or ram-pressure stripped from MaNGA 1-24145, or an extremely low surface brightness (LSB) galaxy. However, the stripping scenario is less favoured according to galaxy merger simulations and the morphology of the \ha~blob. With the current data, we can not distinguish whether this \ha~blob is ejected gas due to a past AGN outburst, or a special category of `ultra-diffuse galaxy' (UDG) interacting with MaNGA 1-24145 that further induces the gas inflow to fuel the AGN in MaNGA 1-24145.

\end{abstract}


\section{INTRODUCTION}
It has been known that the observed number density of satellite galaxies in the Local Group and our own Milky Way is orders of magnitude lower than the predictions from the cosmological simulations, the so-called `missing satellite problem' \citep{kly99,moo99}. One of the popular solutions to this problem is that low-mass halos fail to form stars efficiently such that  they are below the detection limit of most imaging surveys. Under this paradigm,  `dark' galaxies are gas-rich galaxies that do not emit sufficient optical light due to their low efficiency in forming
stars, but are thought to be building blocks of normal star-forming galaxies. This type of galaxy is an
ideal laboratory to study early stages of star formation, which could lead to understanding how
the star formation is triggered. 

It is in general very difficult to identify dark galaxies since they
are extremely faint in the optical. Very few dark candidates have been identified and confirmed to date. The ALFALFA (Arecibo Legacy Fast ALFA) HI survey \citep{gio05} has discovered on the order of $\sim 200$ HI sources not associated with apparent optical counterparts, but most of them are likely to have tidal origins \citep{hay11}. A pilot study suggested that none of remaining objects they have explored is a dark galaxy after cross-checking with data in other wavelengths \citep{can15}.

Recently, \citet{van15a} have identified a new class of low surface brightness (LSB) dwarf galaxies in the Coma cluster, often referred to as `ultra-diffuse' galaxies (UDGs), using the Dragonfly Telescope
Array \citep{abr14}. These UDGs not only have a low surface brightness (24 -- 26 mag arcsec$^{-2}$), but also show extended structures with size comparable to $L^{*}$ galaxies. Although these UDGs are found in special environments and may not necessarily represent the global dwarf population, the discovery of this type of galaxy suggests we might have missed numerous faint dwarfs due to observation limitations in the past.

In addition to the aforementioned efforts, the optical Integral Field Unit (IFU) observations open a plausible window to probe the dwarf populations. Normally, emission lines ionized by star-forming regions, AGNs, or shocks are stronger than the stellar continuum and hence can be easily detected with reasonable integration time. With the large area covered by IFU, the structures of ionized gas can be probed out to several tens of kpc. Isolated ionized gases that are separate from a nearby galaxy or a quasar have been readily studied in IFU as well as spectroscopic observations \citep{fu07a,fu07b,fu08,lin09,hus10,kee12,che16b}. Although the majority of those ionized gas are suggested or inferred to have external origins, such as the result of gas accretion or minor mergers, their nature remains unknown.

Here we report the discovery of a giant \ha~blob which does not have any optical counterpart in deep CFHT $gri$ images. However, the morphology and emission line analyses suggest that it could either be ejected gas due to past AGN activity or a special type of UDGs (or `dark' galaxies). 
In \S2, we describe the multi-wavelength data for this system. We present the main results in \S3. Section 4 discusses the plausible origins of this system and the important implications of our results. Conclusions are given in \S5. Throughout this paper we adopt the following cosmology: \textit{H}$_0$ = 100$h$~\kms Mpc$^{-1}$, $\Omega_{\rm m} =
0.3$ and $\Omega_{\Lambda } = 0.7$. We use a Salpeter IMF when deriving the star formation rate from various observables.
We adopt the Hubble constant $h$ = 0.7 when calculating rest-frame magnitudes. All magnitudes are given in the AB system.

\section{DATA \label{sec:data}}
\subsection{Optical integral field data}
This system, MaNGA 1-24145 ($z$ = 0.0322, RA = 258.84693, DEC =57.43288, M$_{*}$ $\sim 10^{11}$ \Msolar\footnote{based on the NASA-Sloan
Atlas catalog: http://www.nsatlas.org
 }), was observed in the first 1392 galaxies, as part of the on-going SDSS-IV/MaNGA survey \citep{bun15,dro15,law16,yan16a,yan16b,sdss16}. MaNGA is an IFU program to survey for 10k nearby galaxies with a spectral resolution varying from R $\sim$ 1400 at 4000 \AA~ to R $\sim$ 2600 at 9000 \AA. The survey uses the BOSS spectrographs (Smee et al. 2013) on the 2.5m Sloan Foundation Telescope (Gunn et al. 2006). The median full width at half maximum (FHWM) of the MaNGA point spread function (PSF) of the datacube is $\sim$ 2.5".
The MaNGA data used were reduced using the MPL-4 version of the MaNGA data reduction pipeline \citep{law16}.
The spectral line fitting is carried out using Pipe3D pipeline\citep{san16a}. The stellar continuum was first modelled with a linear combination of 12 single stellar population (SSP) templates that were extracted
from the MILES project \citep{san06,vaz10,fal11}. The best-fit stellar continuum is then subtracted from the reduced data spectrum for the emission line measurements. Details of the fitting procedures are described in \citet{san16b}. To ensure reliable measurements, we restrict our analysis to spaxels where the error-to-flux ratio of the line fitting is less than 1 in subsequent analyses.

To correct for dust reddening, we follow the method described in the Appendix of \citet{vog13} to compute the reddening correction by using the Balmer decrement at each spaxel of the IFU cube. An extinction law with $Rv = 4.5$ \citep{fis05} is used. 
The star formation rate (SFR) is then estimated based on this extinction corrected \ha~flux.

\begin{figure*}
\includegraphics[angle=-90,width=17cm]{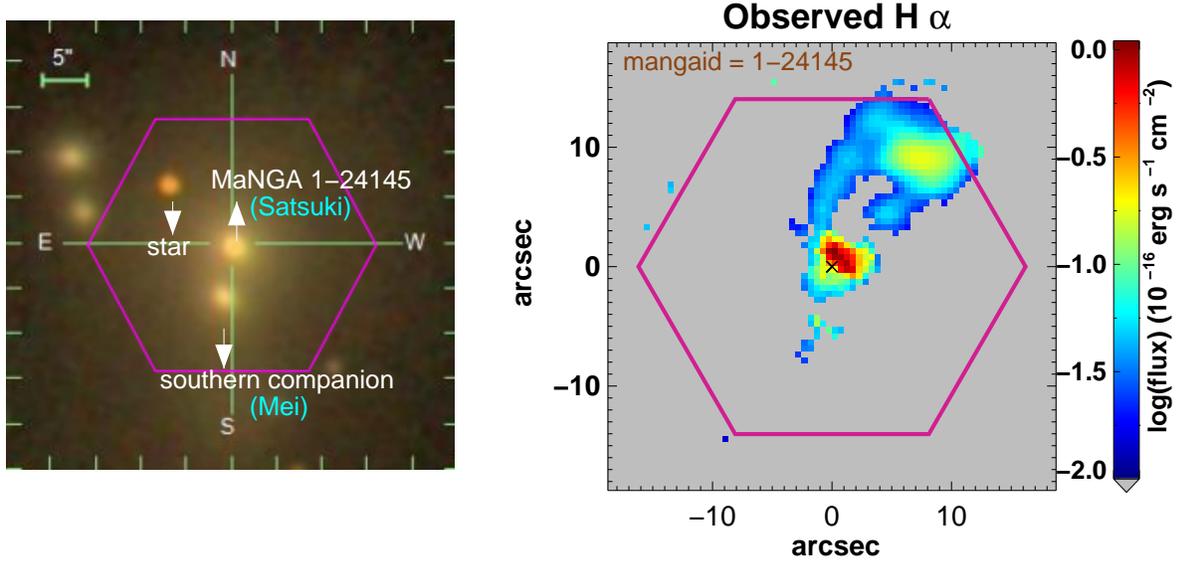}
\caption{Left: The SDSS $gri$ composite image of MaNGA 1-24145 with the MaNGA hexagonal bundle field of view (FoV) overlaid. This system was observed with the 127 fibre bundle of MaNGA, so this hexagon is $\sim$32.5" in diameter. The data extend to regions a bit outside the hexagon because of the dithering. Three distinct objects are visible within the bundle, including two elliptical galaxies (Satsuki and Mei) and one foreground star. Right: the observed \ha~flux map from the MaNGA observations. \label{fig:ha}.}
\end{figure*}

\begin{figure}
\includegraphics[angle=0,width=8.5cm]{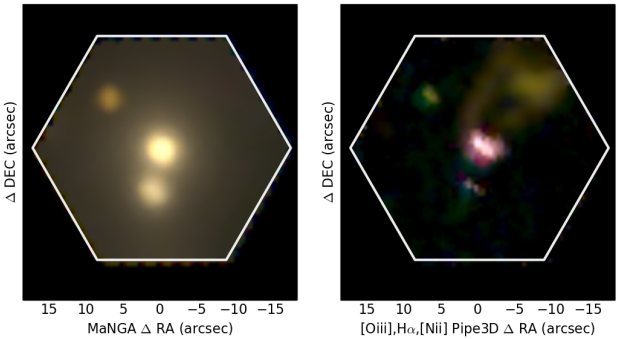}
\caption{Left: The $ugi$ composite image reconstructed using the MaNGA continuum with the MaNGA hexagonal bundle field of view (FoV) overlaid. Right: The [OIII]+\ha+[NII] composite image. The flux scales of the three lines are adjusted in order to highlight the \ha~blob. \label{fig:pipe3Dimage}}
\end{figure}

\begin{figure*}
\includegraphics[angle=-90,width=17cm]{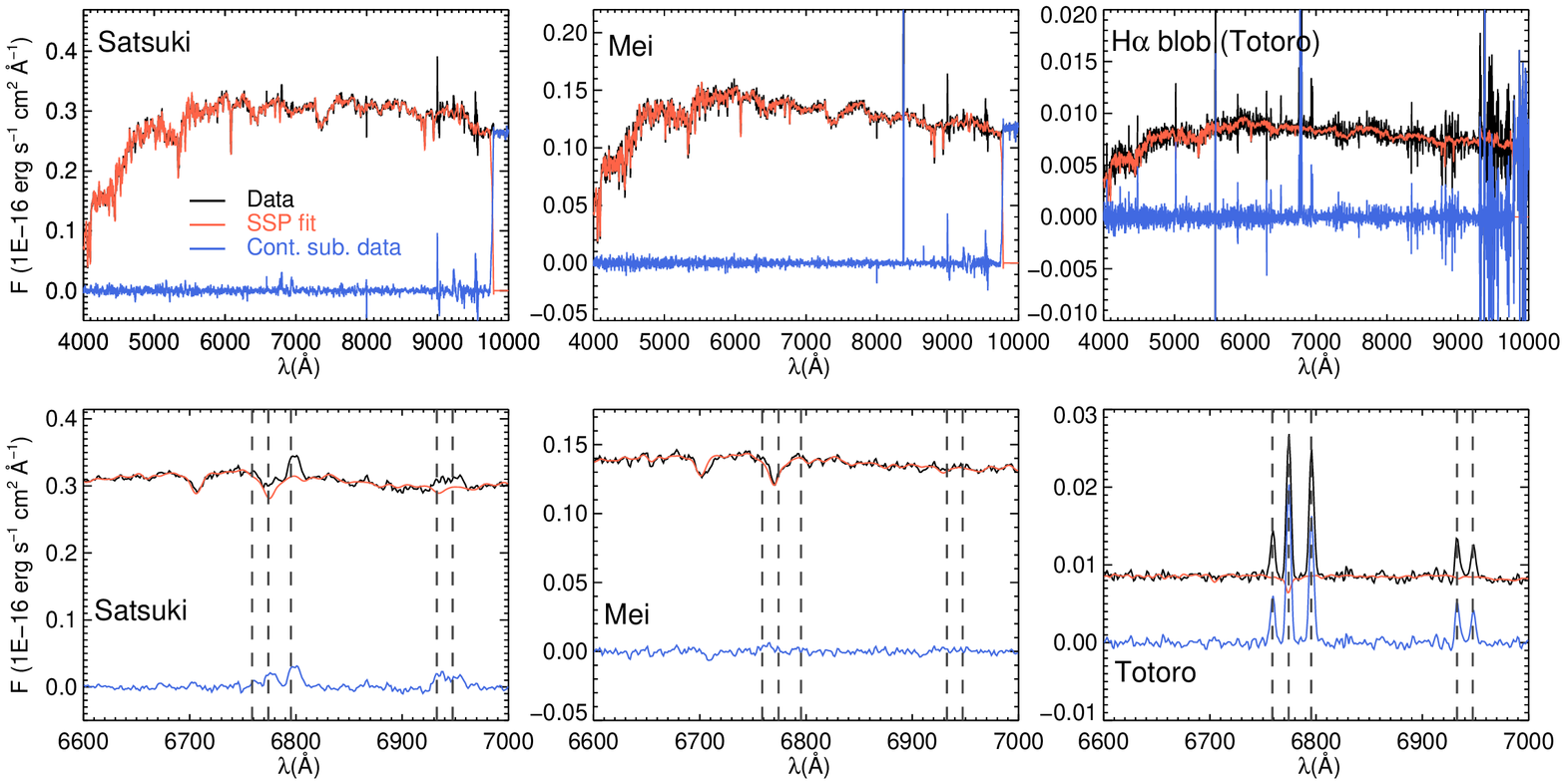}
\caption{Upper panels: The black curves represent the MaNGA spectra of the central spaxel of the main galaxy `Satsuki' (left), the southern companion `Mei' (middle), and the \ha~blob `Totoro' (right). The red curves are the best-fitted SSP model spectra for the stellar continuum. The blue curves show the residual spectra that are used for the emission line fitting. Lower panels: The zoom-in spectra around the \ha~line. The five dashed line denote the observed wavelengths of the \ntwo, \ha, \ntwo, [SII] 6717, and [SII] 6731 lines, respectively. 
\label{fig:spectra}}
\end{figure*}
Figure \ref{fig:ha} shows the SDSS $gri$ composite image (left panel), and the \ha~flux map (right panel) of this system (MaNGA 1-24145
). The SDSS image indicates that MaNGA 1-24145 (nicknamed `Satsuki')
 has a companion galaxy (nicknamed `Mei') located in the lower left to MaNGA 1-24145. 
 The stellar absorption lines suggest that these two galaxies are at similar redshifts with the line-of-sight velocities differed by $\sim$ 200 \kms. In addition, both galaxies are round and red according to the SDSS images. Therefore, these two galaxies likely form a dry (gas-poor) merger system. The compact source in the upper-left corner of the image is a foreground star according to the MaNGA spectrum, and hence it is irrelevant to the dry merger system we are probing for this work.

What is striking about this system is that in the upper-right corner of the \ha~map, there exists a giant \ha~blob (RA = 258.84314; DEC = 57.43529) with which however does not have any optical counterpart in the SDSS images. The size of the \ha~blob is $\sim3.2$ kpc in radius. This \ha~blob (nicknamed `Totoro') is 7.7 kpc away from Satsuki, and is connected to the \ha~emission of Satsuki through tail-like structures. To ensure that the \ha~emission at the position of Totoro is not due to artifacts in the data, we check the MaNGA spectra at various spaxels in the region of Totoro. We found that multiple emission lines, including \ha, \ntwo, \stwo, and \othree~ lines are clearly detected in those regions, suggesting that the \ha~blob is a real feature. Figure \ref{fig:pipe3Dimage} displays the reconstructed $ugr$ image from the MaNGA continuum (left panel) and the [OIII] + \ha + [NII] composite image (right panel). It can be seen that the reconstructed continuum map from the MaNGA data is as smooth as SDSS. The three upper panels of Figure \ref{fig:spectra} show the MaNGA spectra of the central pixels for Satsuki, Mei, and Totoro, respectively. The two elliptical galaxies are mainly composed of old stellar populations, consistent with the morphology classification of being early-type galaxies, whereas the blob is dominated by emission lines. 

\subsection{Optical imaging data \label{sec:optical}}
The imaging data come from two sources: the DR12 release of the SDSS photometric survey \citep{yor00}, which reaches $r\sim22$, and deeper observations with CFHT/MegaCam in $g$, $r$ and $i$. The latter combines archival data downloaded from the CADC server and the data taken in 2015 summer (PI: Lihwai Lin) with the Director Discretionary time (DDT) program. All the MegaCam data were processed and stacked via MegaPipe \citep{gwy08}. The final images have 5$\sigma$ limiting mag of 25.7, 26.2, and 25.2 mag (1" aperture in radius) and surface brightness 5$\sigma$ limit of 26.4, 26.9, and 25.9 mag arcsec$^{-2}$ in $g$, $r$, and $i$, respectively.

\subsection{Radio Continuum}
 
We observed this system with the Karl G. Jansky Very Large Array
(VLA) in the A configuration on 2015 August 20 using the C-band
receiver tuned to $4-6$ GHz ($\lambda = 7.5-5.0$\,cm). The on-source
time is 42 minutes; we observed 3C147 for flux and bandpass calibrations,
and J0920+4441 for phase calibration. Data reduction was carried out
with CASA \citep{McMullin07} using the following steps: (1) standard
calibration using the VLA Data Reduction Pipeline (Chandler et al., in
prep); (2) removal of any portions of the data corrupted by strong
radio frequency interference; and (3) imaging with the task {\tt
CLEAN}. The imaging parameters are the following: MT-MFS deconvolver
with nterms of 2, $0\farcs06$ pixel size, and Briggs weighting with
robust parameter of 0.5. The final image has a $0\farcs39 \times
0\farcs35$ synthesized beam and rms noise at the pointing center of 7
$\mu$Jy beam$^{-1}$.

The radio flux of this source is about 37 $\pm$ 13 $\mu$Jy at 5 GHz.
Assuming that this emission is synchrotron-dominated and so following
a power law $S \propto \nu^{−\alpha}$ with a spectral index $\alpha$ = - 0.7, we derive the
radio luminosity of Satsuki to be 2.2 $\times$ 10$^{20}$ WHz$^{−1}$ at 1.4
GHz. This luminosity implies a star-formation rate of 0.12 \Msolar yr$^{-1}$
using the \citet{bel03} radio SFR indicator, converted to the Salpeter IMF, which is a factor of 2.5 greater than the limit implied by
\ha~emission at the location of the radio source (see Table \ref{tab:property}). Therefore, it is
likely that this radio point source is a faint AGN.

\subsection{HI}
This source was observed as part of the HI-MaNGA programme at the Robert C. Byrd Green Bank Telescope (GBT), which is obtaining HI 21cm observations of a large sample of MaNGA galaxies (AGBT16A\_095, PI: K. Masters). This target was observed on 2016 Feb. 5 for 3 sets of 5 min ON/OFF pairs using the VEGAS spectrometer with a bandwidth of 23.44 MHz, centred on the frequency of 21cm emission redshifted to cz=9653 km~s$^{-1}$. At this frequency the FWHM of the GBT beam is 9\arcmin. No HI emission was detected in this volume to a rms of 1.58 mJy (after smoothing to 5.15 km~s$^{-1}$ velocity resolution). Assuming a velocity spread of 100-400 \kms this non-detection sets an 1-$\sigma$ upper limit of 8.9--9.2 $\times 10^{8}$ \Msolar~for the HI mass of this system \footnote{It is intended that HI-MaNGA data will be released as an SDSS Value Added Catalogue in a future data release from SDSS. In addition the raw data will be publicly available via the NRAO Data Archive at https://archive.nrao.edu a year following observations.}.

\subsection{X-ray}
Satsuki is located within the field of view $\sim$ 47 ks \textit{Chandra} ACIS observation OBSID 4194 (PI: Trevor Ponman), which occurred on 2003 September 17. This pointing of this observation was centered on nearby galaxy NGC 6338 (258.84256, +57.407) [J2000], $\sim$ 2' South of the MaNGA source of interest. Before analysis, this dataset was reprocessed using the Chandra$\_$repro task in CIAO v4.8 \citep{fru06} using CALDB v4.6.3. 

\begin{figure}
\includegraphics[angle=0,width=8.5cm]{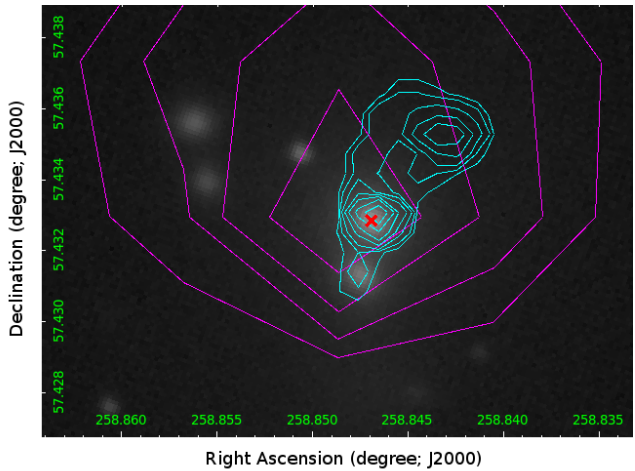}
\caption{The Chandra X-ray contours (magenta) and extinction-corrected MaNGA \ha~contours (cyan) overlaid on the SDSS $r$-band image (background image) of MaNGA 1-24145. The red cross marks the position of the VLA point-source detection. The X-ray contours correspond to 44\%, 30\%, 21\%, and 15\% of the peak value, respectively, whereas the \ha~contours correspond to 100\%, 60\%, 37\%, 23\%, 14\%, 8\%, 5\%, 3\%, and 2\% of the peak value, respectively.
 \label{fig:JVLA}}
\end{figure}

The exposure corrected image of this field (Figure \ref{fig:JVLA}) was generated using the `fluximage' command, and indicates diffuse X-ray emission coincident with the dry merger system, as also shown in \citet{pan12}, which performed a study on a nearby BCG based on the same X-ray dataset. The spectrum of the X-ray emission coincident with the \ha~blob of interest was then extracted using the CIAO script specextract using a source region of a 22" $\times$ 18" ellipse centered at 17:15:23.7, +57.26:05 which encompasses all of the emission. We fit this X-ray spectrum with a single absorbed APEC model -- the emission spectrum from a collisionally-ionized diffuse gas -- assuming the redshift $z$ = 0.032202 measured from optical spectroscopy using XSPEC v12.8.2e \citep{arn96}. This fit results in a reasonable reduced $\chi^{2}$ (1.45 for 88 d.o.f), though somewhat under-predicts the flux $>$ 5 keV. As we will mention in Section 3.5, this dry merger system is part of a small group. The derived X-ray temperature is 1.26$\pm$0.06 keV, consistent with the temperature on group scale \citep{ket13}. No point-like source is found within Satsuki or Totoro, indicating that there is no strong X-ray AGN present in this system.

\section{RESULTS}

\subsection{The optical morphology \label{sec:obs}}

\begin{figure*}
\includegraphics[angle=0,width=17cm]{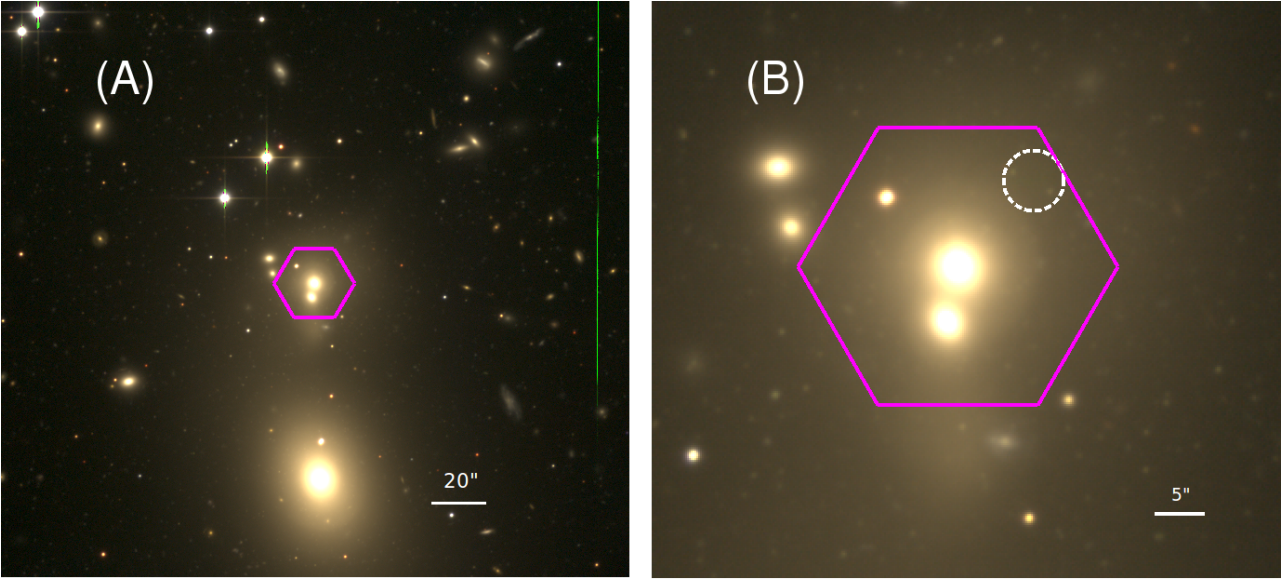}
\caption{
(A) CFHT $gri$ composite color image for MaNGA 1-24145 with the MaNGA hexagonal FoV overlaid. A bright BCG is $\sim$ 43 kpc away to the South. (B) A zoom-in picture of (A).
The white circle marks the location of Totoro. In both panels, North is up and East is left. \label{fig:cfht}}
\end{figure*}

To ensure that the absence of the optical counterpart at the position of Totoro as shown in Figure \ref{fig:ha} is not due to relatively shallow depth of SDSS imaging, we carried out a follow-up observation for this system with CFHT/MegaCam and combined it with the archival data (see section \ref{sec:optical}). Figure \ref{fig:cfht} displays the $gri$ composite image of this galaxy. It is clear that there are extended stellar halos surrounding the two galaxies. Again at the position of the \ha~blob, no apparent optical continuum is revealed (see the right panel of Figure \ref{fig:cfht}). 

\begin{figure}
\includegraphics[angle=0,width=8cm]{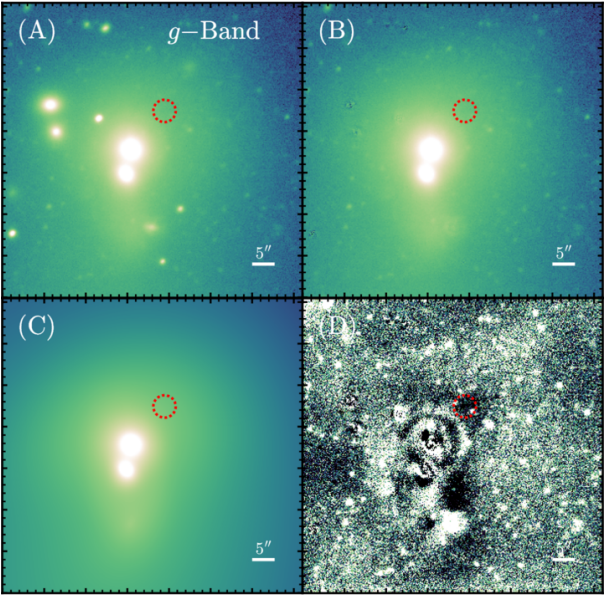}
\caption{
(A) The original CFHT Megacam $g-$band image, (B) the $g-$band image after subtracting nearby satellite galaxies, (C) model image for (B), and (D) the residual images produced by GALFIT. The dashed while circles mark the position of Totoro. The white bars in the lower-right corner of each panel corresponds to a scale of 5". \label{fig:song}}
\end{figure}

\begin{figure*}
\includegraphics[angle=-90,width=17cm]{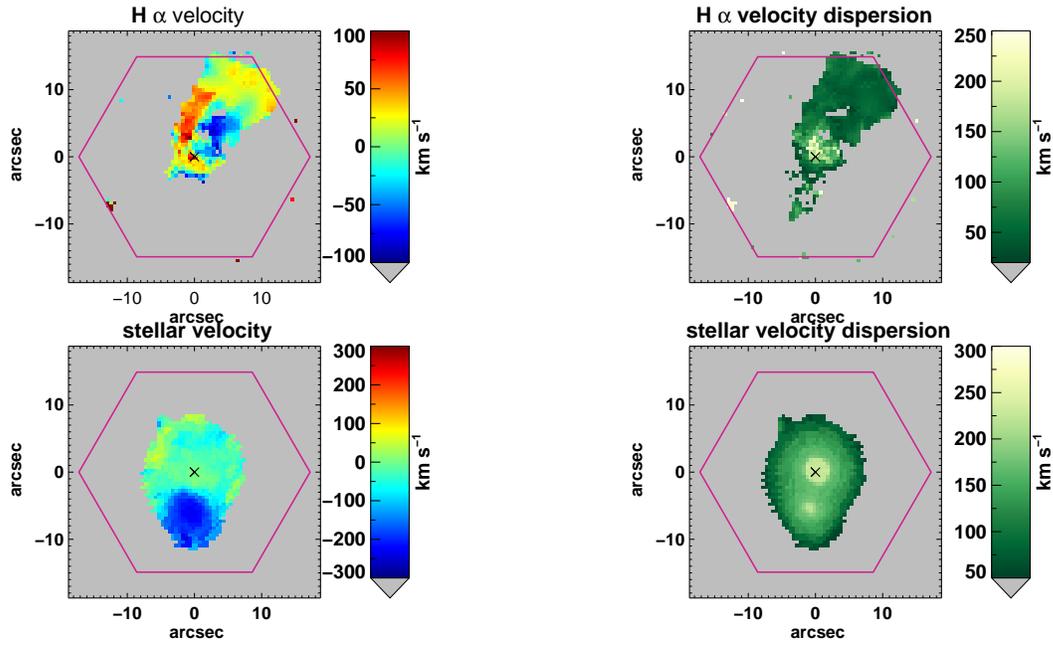}
\caption{
Velocity fields of MaNGA 1-24145
 based on \ha~line (upper panels) and stellar components (lower panels). The left panels show the radial velocities and the right panels show the velocity dispersions. \label{fig:velocity}}
\end{figure*}

\begin{figure*}
\includegraphics[angle=-90,width=17cm]{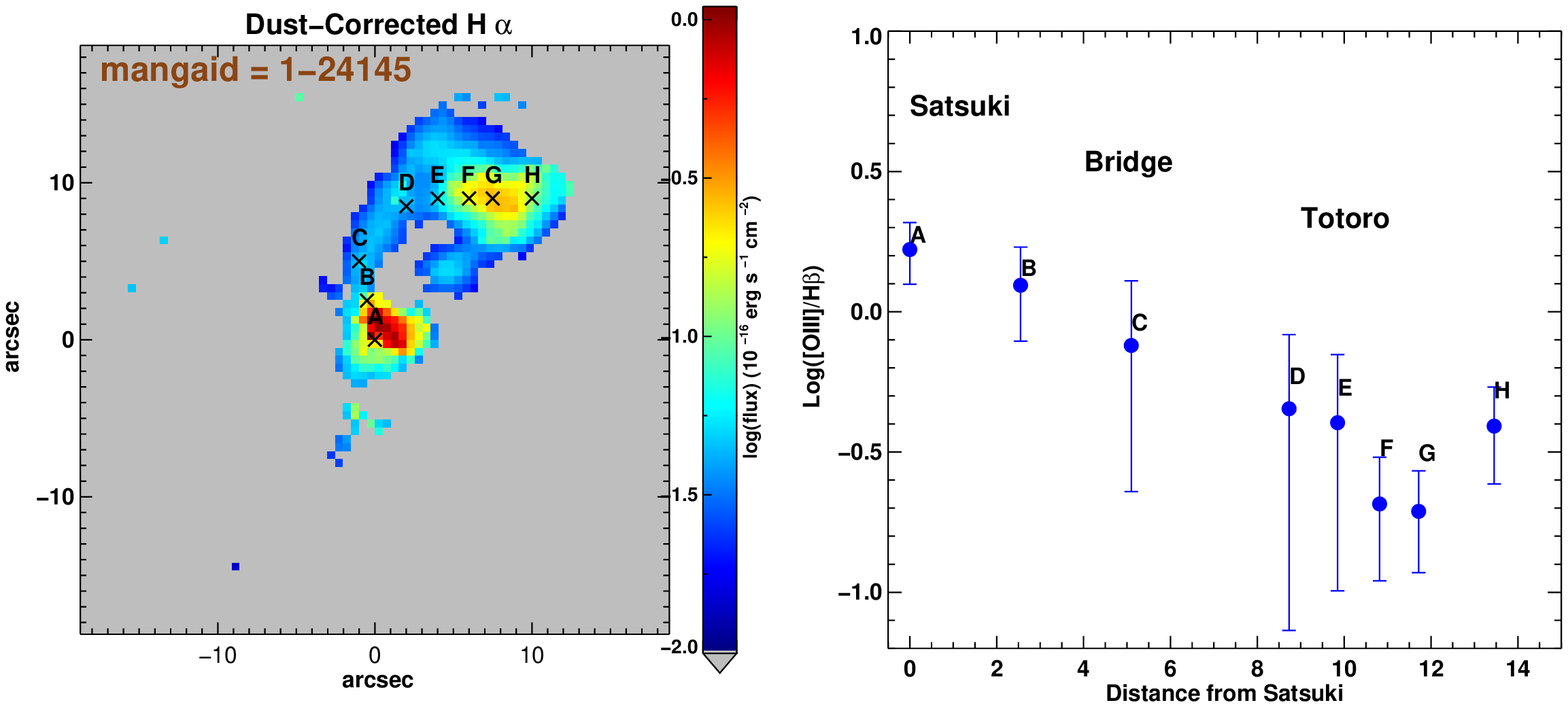}
\caption{Left: The extinction-corrected \ha~map of MaNGA 1-24145
. Right: The \othree/\hb~ratio as a function of distance from the center of Satsuki
. The letters (A,B,C,..etc) mark different locations of this system.
\label{fig:o3hb}}
\end{figure*}

To more clearly see the underlying surface brightness structure at the location
of Totoro, we build detailed photometric models for this merging system
using \texttt{GALFIT\ v3.0.2} (Peng\ et al.\ 2002, 2010).The MecaCam $g$-band image is used here for this purpose because it may better reveal the continuum of Totoro if has had recent star formation (see \S 4). Since we are not interested in
the overall structures of this complex merging system, we will only focus on a
400$\times$400 pixels region centered on the MaNGA bundle.  We
construct a mask image to exclude isolated objects around the merging system from the
fitting.  The PSF of the image was modelled using the \texttt{SExtractor} and 
\texttt{PSFex} routines, and the PSF image used in the modelling was extracted using the 
central coordinate of the MaNGA bundle.  Meanwhile, on top of the extended evelope, there
are 11 objects (most are galaxies) that can not be easily masked out.   We model them 
separately, then subtract them from the image.  All of these smaller objects can be 
well modelled by a single- or double-\ser model locally with the help of an additional 
\ser and sky components to account for the envelope in the background.  As shown in the 
panel (B) of Figure \ref{fig:song}, they have been removed smoothly from the input image without any 
significant residual pattern.  Using this ``cleaned'' image as input, we model the two 
merging galaxies along with their ``common envelope'' together using different
combinations of \ser components.  We started with three \ser components (one for each 
galaxy, additional one for the envelope), and gradually build up the complexity by adding 
more \ser component.  As we simply want to achieve smooth residual map to study the 
underlying structure, the number of the \ser components and the detailed parameters are
not a concern as long as each component behaves normally (e.g. Gu\ et al.\ 2013).  
For each object (including the envelope), all \ser components are constrained to have the
same center, and only symmetric \ser components are used here.  After visualizing the
residual of initial model, it becomes clear that, on the lower part of the image, there is
an additional surface brightness enhancement (caused by the merging process) that are not well fit by the simple model.  Eventually, the best model we achieved include 7 \ser components plus a
tilted-plane sky background component.  Both of the merging galaxies, and the extended 
envelope are described by 2 \ser components; while the surface brightness in the south is 
modelled using single \ser component.  All components behave regularly in term of size and 
shape, except for the central component for the brighter galaxy, no component has \ser 
index larger than 2.0.  

The panels (C) and (D) of Figure \ref{fig:song} show the model image and the residual maps of this
model reconstruction, respectively. It clearly reveals a rich system of shells and tidal tails around the main MaNGA 
galaxy, indicating an on-going interaction between these two galaxies.  Although the
residual map is not perfectly smooth, no optical counterpart is found at the position of
Totoro.  

Reducing the number of components used, or changing the initial guess will not affect the
above conclusion.  Adding more components could not further improve the residual map,
but results in ill-behaved \ser components.  Given the complex nature of this merging
system, we also try to invoke the asymmetric features in \texttt{GALFIT}, especially the 
1st (global lopsidedness) and 4th (boxiness of the isophote) Fourier components (see 
Peng\ et al.\ 2010 for details).  The residual map shows improvements around the tidal
features, but does not change the conclusion that there is no apparent optical counterpart
for Totoro \footnote{The results still hold if we repeat the analysis using the $i$-band image, which is more sensitive to the stellar mass distributions.}.

\subsection{Kinematics from the MaNGA Observations\label{sec:velocity}}
Figure \ref{fig:velocity} displays the velocity and dispersion maps for this system. 
While the stellar velocity field indicates that the main galaxy Satsuki
 is primarily pressure supported, the gas component reveals more complex structures. The central galaxy shows a weak rotation structure, while there is strong variation in the line-of-sight velocity across Totoro region, redshifted in the left tail and blueshifted in the right tail.  The inconsistent velocity fields between stars and gas suggests that some part of the gas of Satsuki might have been either accreted or ejected recently. In the former case, it is similar to those early-type galaxies 
that exhibit misaligned gas and stellar kinematics, the so-called early-type `counter rotators' \citep{sar06,dav11,che16,jin16}, although counter rotators
in general are defined for systems with a rotating stellar component.  The gas inflow scenario is also consistent with the \ha~morphology which shows bridge (tail)-like structures that connect Totoro and Satsuki. On the other hand, the complicated velocity field can also be explained if the main galaxy Satsuki underwent a strong outburst phase, during which Totoro was expelled from Satuski.

\subsection{Excitation State \label{sec:bpt}}

In addition to \ha~and two \ntwo~lines, some other weak lines such as \stwo~ and \othree~ are also detected in both Satsuki and Totoro, allowing us to 
probe the ionization state for this system. Figure \ref{fig:o3hb} shows the \othree/\hb~ ratio, one of the frequently used ionization parameters, as a function of the distance from the main galaxy. There exists a strong \othree/\hb~gradient, decreasing from the main galaxy Satsuki (location A) to the left bridge (locations B, C, and D) that connects to Totoro (location E, F, G, H, and I). On the other hand, the \othree/\hb~ ratio is nearly constant across Totoro.

\begin{figure*}
\includegraphics[angle=-90,width=17cm]{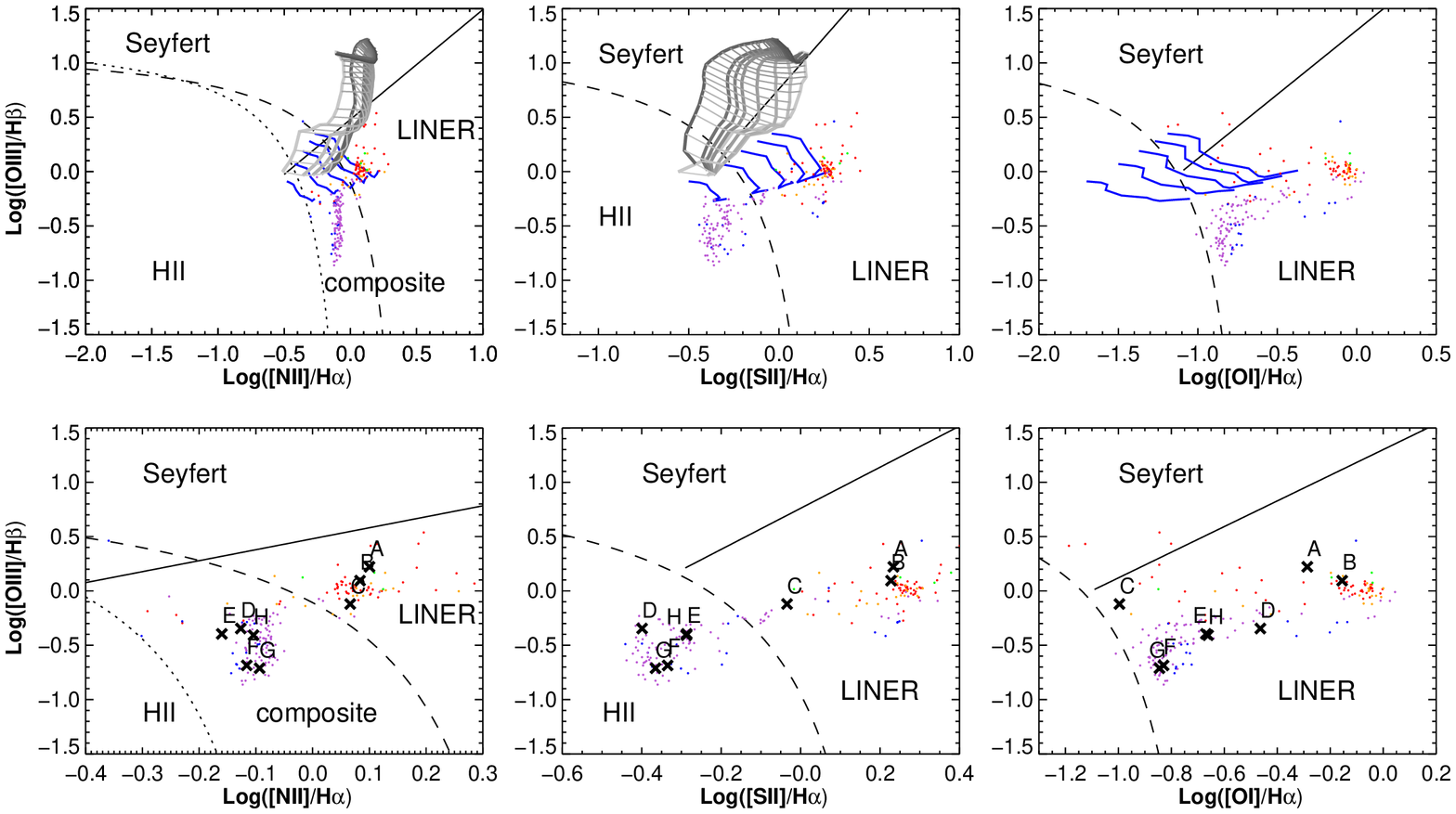}
\caption{Upper panels: BPT diagnostic diagrams for MaNGA 1-24145
 based on the \othree/\hb~ vs. \ntwo/\ha~ (left panels), \othree/\hb~ vs. \stwo/\ha~(middle panels), and \othree/\hb~ vs. \oone/\ha~(right panels). Various colors on the data points indicate the physical separation from the center of Satsuki (from near to far: red, origin, green, blue, purple). The solid, dashed, and dotted lines show the classification lines suggested by \citet{cid10}, \citet{kew01}, and \citet{kau03}, respectively. The blue curves display the model predictions of the shock and photoionization mixing sequences by \citet{ho14}. Each line corresponds to a certain shock fraction (from bottom to top: 20\% to 100\%). The shock velocity ranges from 100 to 300 \kms (from left to right). The MAPPINGS III shock+precursor models (gray grid) with $n$ = 1 cm$^{-3}$ from \citet{all08} are also shown for comparison. The thick lines represent constant magnetic parameter while the thin lines display the constant shock velocity ranging from 200 to 1000 \kms (bottom to top; with 25 \kms intervals).
Bottom panels: A zoomed-in view of the upper panels. The letters (A,B,C,..etc) mark different locations of this system, following the definitions of Figure \ref{fig:o3hb}.
\label{fig:bpt}}
\end{figure*}

\begin{figure*}
\includegraphics[angle=-90,width=17cm]{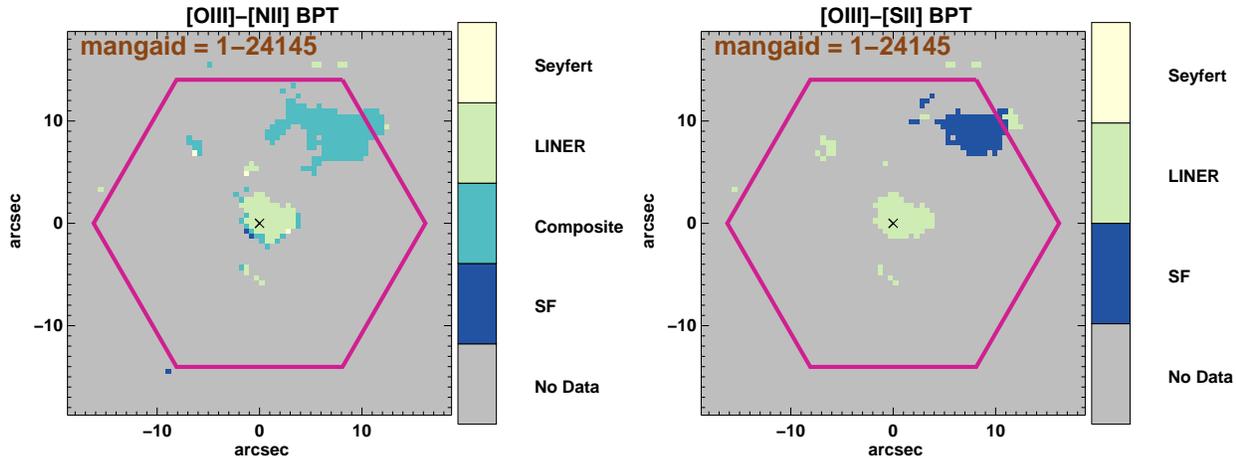}
\caption{BPT classification maps for MaNGA 1-24145
 based on the \othree/\hb~ vs. \ntwo/\ha~ (left panel) and \othree/\hb~ and \stwo/\ha~(right panel). The [OI] diagnostic map is not included here as all spaxels are classified as LINER.
\label{fig:bptmap}}
\end{figure*}

The multiple line detections also allow us to classify the emission line regions into HII or AGN regimes using the standard Baldwin-Phillips-Terlevich (BPT Baldwin, Phillips \& Terlevich 1981; Veilleux \& Osterbrock 1987; Kauffmann et al. 2003; Kewley et al. 2006) excitation diagnostic diagrams. Here we apply three types of line diagnostics based on four line ratios, \othree/\hb, \ntwo/\ha~, \stwo/\ha, and \oone/\ha. Figures \ref{fig:bpt} and \ref{fig:bptmap} display the line ratio diagrams  and classification maps for this system, respectively. We adopt the dividing curves suggested in the literature \citep[e.g.,][]{kew01,kau03,cid10} to separate various regions. All the three classifications indicate `LINER'-type excitations for Satsuki. As the LINER emission is extended across the entire main galaxy, this object falls in to the extended LIER (eLIER) category according to the classification scheme by \citet{bel16a,bel16b}

On the other hand, in the regions of Totoro, the line ratios are consistent with the `composite', `HI'. and 'LINER' regions when using the [NII], [SII], and [OI] diagnostics, respectively. In the remaining part of this paper, we treat Totoro as `composite' regions based on the [NII], as it allows for the intermediate excitation state as opposed to the other two methods.

\begin{figure}
\includegraphics[angle=-90,width=11cm]{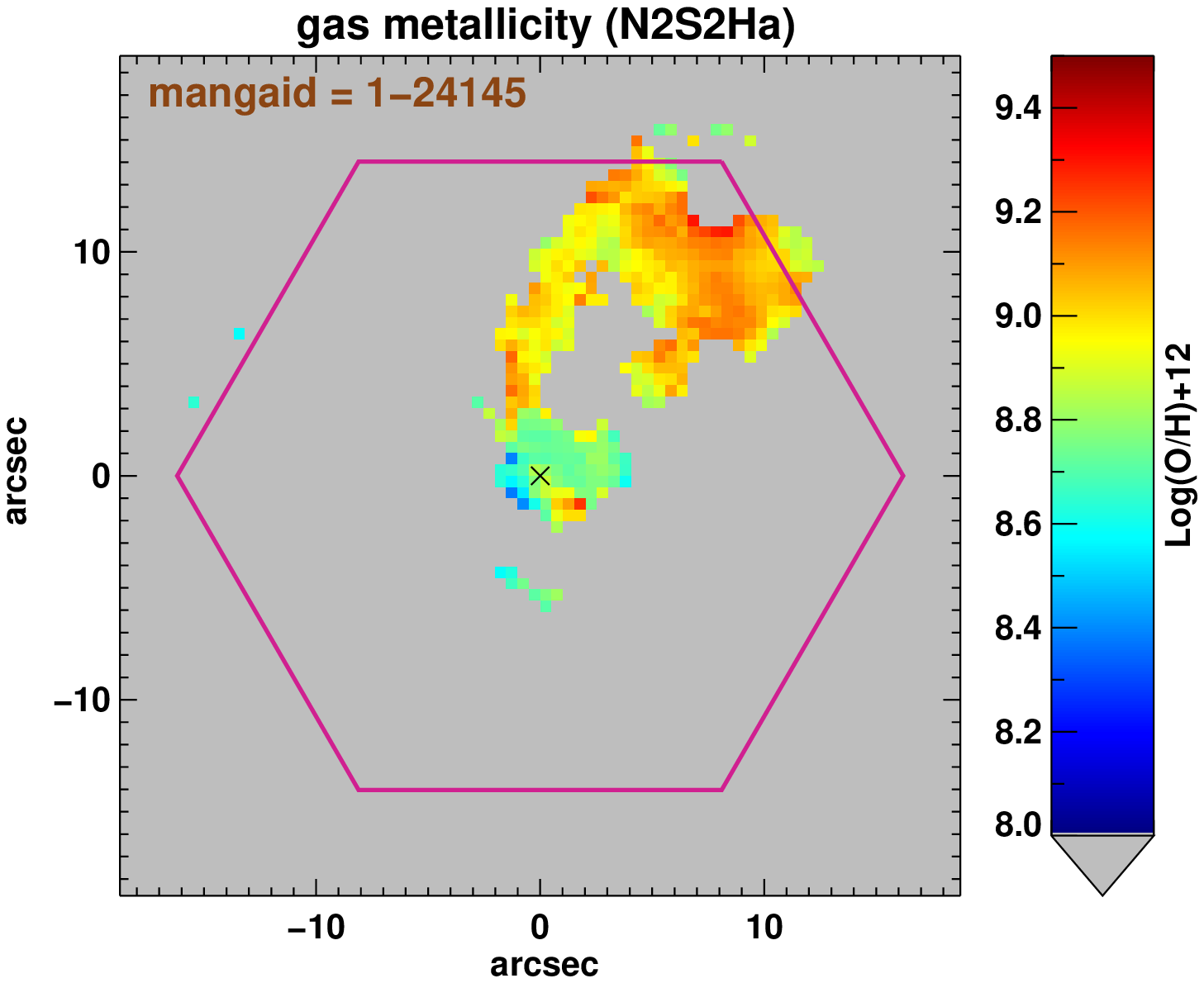}
\caption{The gas metallcity map of MaNGA 1-24145
, computed using the method described in \citet{dop16} based on the emission line ratios among [NII], [SII], and \ha.
\label{fig:metal}}
\end{figure}

\subsection{Gas metallicity \label{sec:metal}}
To further understand the properties of Totoro, we also measure the gas-phase metallicity (Z) for this system. Conventionally, there are many ways to estimate the gas metallicity through emission line ratios \citep[see][]{kew08}. However, most of those tracers are calibrated against the HII regions, and may not be applicable to systems with ionization parameters or interstellar medium (ISM) pressure different from typical HII regions. Here, we adopt the `N2S2\ha'~ calibrator that is suggested to be less sensitive to the ionization parameters \citep{dop16} to estimate the metallicity. The N2S2\ha~calibration can be expressed as the following:

\begin{equation}\label{eq:N2S2ha}
12 + \mathrm{log (O/H)} = 8.77 + \mathrm{N2S2H}\alpha
\end{equation}
where N2S2\ha = Log([NII]/[SII]) + 0.264Log([NII]/\ha). The [NII] to [SII] ratio is found to nearly independent of the AGN luminosity and hence provides a good metallicity indicator for even AGNs \citep{ste13}. However, we note with caution that the derived metallicity may still be subject to large systematic uncertainty since the line ratios of the main galaxy and Totoro are located in the LINER and composite regions, respectively. 

Figure \ref{fig:metal} shows the Z(N2S2\ha) map for this system. The metallicity of the gas around Satsuki is close to solar and is $\sim$ 0.3 dex lower than the expected value (Z = 9.1) for a massive galaxy with a stellar mass of $10^{11}$ \Msolar~  based on the local mass--metallicity relation \citep{tre04}. The offset we see here, however, may be easily accounted for by the different metallicity calibrators adopted. On the other hand, the averaged metallicity in the \ha~ region is greater for Totoro than Satsuki by a factor of 0.3 dex and is more consistent with the gas metalltcity of high mass galaxies. Although the difference is significantly larger than the statistical uncertainty (0.03-0.1 dex) in the metallicity measurement, it is difficult to interpret this result given the systematic uncertainty in the metallicity measurement due to their different ionization states.

\subsection{Environment \label{sec:environment}}
The environment of this system
 is rather complex. As already mentioned, it has an early-type companion (Mei) just 4 kpc away to the South, which makes them a possible dry merger candidate. Moreover, Satsuki
 is also part of the system MCG+10+24-117, a small group falling onto a galaxy cluster with NGC 6338 as a central Brightest Cluster Galaxy that is 43 kpc away to the south (see Figure \ref{fig:cfht}; Pandge et al. 2012; Dupke et al. 2013).

\section{DISCUSSION}\label{sec:discussion}

\subsection{The origin of the \ha~ blob}
Using the first-year MaNGA data, we discover a giant \ha~blob, Totoro, associated with a dry merger system. This object, however, does not have any optical counterparts down to 26.9 mag arcsec$^{-2}$ in deep CFHT/MegaCam $g,r$, and $i$ images. 
There are several possible scenarios to explain the origins of Totoro:

Scenario 1: Totoro is associated with gas tidally stripped from Satsuki during
the interaction between Satsuki and Mei. 

Scenario 2: Totoro is associated with the gas ram-pressure stripped during the infall of
Satsuki
 toward the center of NGC 6338 galaxy cluster, similar to NGC4569 located in the Virgo cluster (Boselli et al. 2016).
 
Scenario 3: Totoro is associated with gas ejected from Satsuki by an AGN outflow during
the mergers between Satsuki and Mei. If the central black hole of
Satsuki is turned on during mergers, the energy can ionize the stripped gas, resulting in
the \ha~emissions, similar to the known `Hanny's voorwerp' phenomenon (Lintott et al. 2009).

Scenario 4: Totoro is an UDG (or alternatively, a LSB galaxy), which falls
below the detection limit (26.9 mag arcsec$^{-2}$ in $r$-band) of CFHT imaging data, making it a `dark' galaxy. The morphology of Totoro (Figure 1), especially the (tail) bridge-like structures that connect between Satsuki and Totoro, indicates that Satsuki
 is likely under interaction with the hosting galaxy of
Totoro, which has a comparable physical size as Satsuki.

To estimate the mass of the warm gas component, we follow the approach adopted by \citet{che16b}. We first estimate the electron density $n_{e}$ to be $\sim$ 260 cm$^{-3}$ based on the median value of the [SII]6717/[SII]6731 ratio in the region of Totoro following Equation 3 of \citet{pro14}, which assumes the electron temperature $T_{e} = 10,000K$. Next, we calculate the extinction-corrected \hb~luminosity, $L_{\mathrm{H}\beta}$, to be 3.1$\times 10^{39}$ erg s$^{-1}$. The ionized gas mass can then  be derived using the following equation \citep{ost06,fu12}:
\begin{equation}\label{eq:mass}
\frac{M_{\rm HII}} {6.8\times 10^{7}~ \rm M_{\odot}} = \bigg(\frac{L_{\mathrm{H}\beta}}{10^{40}~\mathrm{erg s^{-1}}}\bigg)\bigg(\frac{n_{e}}{1~  \mathrm{cm}^{-3}}\bigg)^{-1}.
\end{equation}
We obtain the ionized gas mass $\sim$ 8.2$\times$ 10$^{4}$ \Msolar.

The HI observation of this system provides an upper limit of (8.9--9.2) $\times 10^{8}$ \Msolar~for the HI mass due to the null detection. We account for He gas by applying a factor of 1.33 to this upper limit. We can also infer the H$_{2}$ content of this Totoro using the \ha~emission. The integrated \ha~flux over spaxels within Totoro region is 3.3$\times 10^{-15}$ erg s$^{-1}$ cm$^{-2}$, corresponding to a luminosity of 7.8$\times 10^{39}$ erg s$^{-1}$. Assuming all the \ha~flux results from star formation, we obtain the total star formation rate (SFR) of Totoro to be 0.059\Msolar yr$^{-1}$ using the conversion between the \ha~luminosity and star formation rate from \citet{ken98}.
We then estimate the total H$_2$ gas mass ( = SFR x t$_{dep}$) of the H$\alpha$ blob to be $\sim1.2\times 10^8\,M_\odot$ assuming a typical gas depletion time (t$_{dep}$) of 2 Gyr. This value is also an upper limit since we have  assumed that all the \ha~fluxes are contributed by the star formation. This implies an upper limit of the entire cold gas (HI + He + H$_{2}$) of $\sim$ 1.3 $\times 10^{9}$ \Msolar.

Although Satsuki and its companion Mei are consistent with being early-type galaxies (ETGs), the amount of the cold gas in Totoro is not totally unexpected if it is part of Satsuki. For example, recent works have found that a significant fraction of massive ETGs that show signs of star formation possess cold gas with gas mass comparable to this Totoro \citep{osu15,dav16}. Therefore, it is reasonable to speculate that Totoro may  originally be part (if not all) of the cold gas contained Satsuki, and then get ejected during the galaxy-galaxy interaction or because of the ram-pressure stripping (scenarios 1 to 3, respectively).

To test the galaxy interaction scenario (scenario 1), we have performed and examined a set of N-body
simulations of the interaction of two Elliptical galaxies (E-E) with the
code Gadget2 (Springel 2005). Different initial orbital configurations
(e.g pericentric distance) have been considered in a similar way to Peirani
et al. (2010). Although we did not include any gas component in our dry merger simulations, the features in the stellar components can still be useful to understand the system as the stellar streams well trace the gas streams before 1 Gyr during galaxy interaction. 

Among all the cases we have looked at, we did not see the formation of any clear stellar stream or a centrally concentrated blob-like structures during the interaction. This strongly suggested that Totoro is unlikely to be produced during an E-E interaction. In addition, based on other existing gas-rich merger simulations (see
for instance Barnes \& Hernquist 1996; Di Matteo et al. 2007; Peirani et
al. 2010), the prominent gas (or stellar) streams are expected to be
formed when at least one of merger progenitors have extended disk structures, unlikely to be the case for this dry merger system even if these two ellipticals have small amount of gas. Furthermore, such latter simulations also suggest that it is very
difficult to produce such blob-like structure. Although tidal dwarf
galaxies can be formed during the interaction of disk galaxies (see for
instance Bournaud, Duc et Masset 2003; Duc, Bournaud et Masset 2004) in these cases, the
blob-like structure is more extended with no clear star formation
activity. Nevertheless, we note that the simulations we have examined have the following caveats: 1) the possible parameter space is extensive, so it is impossible to fully explore, and 2) the stellar/AGN feedback which could drive galactic scale outflows and these physics are not included. Such advanced simulations are beyond the scope of this work. We defer more detailed comparisons with simulations to a future work.

On the other hand, the centrally concentrated blob-like structure is not expected in the ram-pressure stripped gas (scenario 2), either, which often shows the tail and/or `jellyfish' like structures (e.g., Boselli et al. 2016). Moreover, galaxies that show ram-pressure stripping phenomena are mostly gas-rich late-type galaxies, unlike Satsuki.

Another explanation for Totoro is the materials ejected by an AGN located in Satsuki (scenario 3), similar to the extended emission-line regions found around quasars \citep{fu07a, fu07b}.  The outflow scenario is also consistent with the complicated velocity field seen in the gas component of this system. On the other hand,
as discussed in \S 2, we do not detect an X-ray point source or an extended radio jet in the main galaxy, suggesting that there is no on-going strong AGN activity. However, it has been known that the AGN brightness can vary over timescales of several to 10$^{5}$ years \citep{den14,mce16}, either due to a change in the black-hole accretion rate or Tidal-Disruption Events \citep[TDEs, e.g.][]{sax12,mer15}. Therefore,  we can not rule out the possibility of a recent past AGN outflow, similar to the notable phenomena of Henny's Voorwerp, in which the ionizing source has already diminished whereas a clump of ionized gas is found several kpc away. Nevertheless, it is unclear why the ejected gas would have a higher metallicity than than the gas remaining in Satsuki (although it is unclear whether the metallicity difference is real given the potential systematics in our metallicity estimate).

An alternative explanation of the \ha~blob is a separate faint galaxy, interacting with the dry merger system (scenario 4). 
A few morphological features, for example, the centrally concentrated \ha~and the bridges extended from Totoro toward Satsuki, are most consistent with Totoro being a separated gas component. According to numerical simulations, these features can indeed be explained by the interaction between the main galaxy and a faint disk galaxy (scenario 1) (see for instance, Peirani et al. 2010; Cheung
et al. 2016).

Recently, a class of UDGs has been identified in the Coma cluster and several low-redshift clusters \citep{van15a,van16}. These UDGs are surprising large in size ($r_{eff}$ = 1.5 -- 4.6 kpc) despite of their low stellar masses ($< 10^{8}$ \Msolar). The majority of these UDGs are found to lie on the red-sequence, indicating old stellar populations and quenched star formation activities \citep{van16}. Possible formation mechanisms include that gas is stripped by the ram pressure when falling into the cluster, which prevents subsequent star formation, or that gas is expelled due to strong stellar or supernova feedback for galaxies with halo mass between $10^{10}$--$10^{11}$ \Msolar~\citep{dicin16}. Nevertheless, it remains an open question regarding the origin of UDGs and whether these UDGs are already quenched before falling into the cluster environments. Totoro identified in this work has a comparable size ($\sim$ 3.2 kpc) as the UDGs. However, the averaged surface brightness of Totoro has an upper limit of 26.9 mag arcsec$^{-2}$, at least 1-3 mag dimmer compared to that of the known UDGS ranging from 24 to 26 mag arcsec$^{-2}$. Another aspect of Totoro that is distinct from the known UDGs is that the latter are generally old in the stellar populations as indicated from their red colors, whereas Totoro shown in this work is likely to have small amount of on-going star formation based on the BPT diagnostics but with little old stellar populations.  Therefore, Totoro may represent a different category of LSB galaxy from the quiescent UDGs.

The non-detection of optical light associated with Totoro can provide an upper limit of the amount of star formation. Assuming that this \ha~cloud is a young star-forming system with age $< 0.1$ Gyr, the flux in the optical regime is expected to be comparable to that in the UV (1500--2800 \AA) in the case of no dust extinction. Following the conversion between the UV luminosity, \ha~luminosity, and the star formation rate given by \citep{ken98} and adopt the surface brightness limit (26.9 mag arcsec$^{-2}$) from our CFHT MegaCam $r-$band observation, we estimate that the corresponding \ha~surface density of this Totoro is on the order of 1.1 $\times10^{-17}$erg s$^{-1}$ cm$^{-1}$ arcsec$^{-2}$, which is one magnitude lower than the peak value of Totoro. By integrating it over the \ha~region, we estimate that the star formation contributes less than 37\% to the total \ha~flux, otherwise we should be able to detect the optical counterpart.

Given the very low amount of star formation that can possibly occur in Totoro, we speculate that Totoro is  a gas cloud that fails to form stars efficiently, and thus a `dark' gas cloud. The origin of this gas cloud, whether it is associated with a satellite dark matter, or a pure gas cloud, is difficult to pin down. However, the latter scenario is unlikely since it would be difficult to maintain the kpc-scale gas cloud against gravity without invoking the existence of a dark matter halo.   If Totoro is indeed hosted by a subhalo, it would be a strong evidence of the existence of `dark' subhalos, which help to alleviate the missing satellite problem. Since there is no stellar component in the region of Totoro, we could roughly estimate the halo mass of Totoro based on the gas velocity dispersion using the virial theorem. Taking R = 3.2 kpc and $\sigma_{gas} = 50$ \kms, we obtain the halo mass M$_{halo}\sim 5.6\times10^{9}$ \Msolar.

On the other hand, the high metallicity of Totoro (see Section 3.4) is unexpected for a typical dwarf galaxy. One possible explanation is that the gas cloud has been enriched by the surrounding environment, and may has a different evolution process from the typical known dwarf galaxies. However, we caution that the metallicity measurement presented in this work is subject to large uncertainty since the ionization source is not well constrained.

\subsection{Sources of ionization}

\begin{deluxetable*}{llllllllll}
\tabletypesize{\scriptsize}
\tablewidth{0pt}
\tablecaption{Properties of MaNGA 1-24145 (Satsuki), its southern companion (Mei), and the \ha~blob (Totoro).\label{tab:property}}
\tablehead{
	\colhead{Object} &
    \colhead{$z$} &
    \colhead{RA} &
    \colhead{DEC} &
    \colhead{M$_{*}$ (\Msolar)} &
    \colhead{M$_{\rm HII}$ (\Msolar)} &
    \colhead{M$_{H_{2}}$ (\Msolar)} &
    \colhead{M$_{HI}$ (\Msolar)} &
    \colhead{M$_{halo}$ (\Msolar)} &
    \colhead{SFR (yr$^{-1}$ \Msolar)} 
    }

\startdata
Satsuki & 0.0322  & 258.84695 & 57.43288 & $1.2 \times 10^{11}$ & \nodata & \nodata & $< 9.2 \times 10^{8}$ & \nodata & $<$ 0.049$^{b}$\\
Mei & 0.0322  & 258.84750 & 57.43133 & $4.0 \times 10^{10~a}$ & \nodata  & \nodata  & $< 9.2 \times 10^{8}$ & \nodata & \nodata\\
Totoro & 0.0322  & 258.84314 & 57.43529 & \nodata & $8.2 \times 10^{4}$  & $< 1.2 \times 10^{8}$ & $< 9.2 \times 10^{8}$ & $5.6 \times 10^{9}$ & $<$ 0.059$^{b}$
\enddata

\tablecomments{$^{(a)}$ This is scaled from the stellar mass of Satsuki by using the difference in their SDSS $r$-band magnitudes.; $^{(b)}$ This upper limit is derived assuming all the \ha~fluxes come from the star formation.}

\end{deluxetable*}

Shocks can often lead to line ratios similar to those occupying the composite regions \citep{ho14}. In the case where there are shocks, the gas velocity dispersion ($\sigma_{gas}$) is expected to be as high as several hundreds of \kms. As revealed in Figure \ref{fig:velocity}, $\sigma_{gas}$ in the region of Totoro is $\sim 50$ \kms. Although it is close to the lower end of velocity dispersion distribution that have been found in typical shocked regions, we can not ruled out the shock excitation of the emission lines. To gain insights to whether shock is responsible for producing the ionized ratios seen in this system, we show the shock and photoionization mixing models from \citet{ho14} as blue curves in the upper panels of Figure \ref{fig:bpt}. These models are produced based on the \texttt{MAPPINGS IV code} \citep{dop13} and span a wide range of shock fractions (from 20\% to 100\%) and shock velocities (from 100 to 300 \kms). As it can be seen, the models predict a greater \othree/\hb~ratio than what is observed in the data and do not cover the majority of the regions occupied by the data points even at shock velocity up to $\sim300$ \kms. In addition to the pure shock models, we also compare the data to the shock + precursor models of Allen et al. (2008) with $n$ = 1 cm$^{-3}$ and solar metallicity, shown as the gray grid in figure \ref{fig:bpt}. Although the grid starts with the shock velocity of 200 \kms, it is expected by extrapolation that models with lower velocity values still can not reproduce the [NII]/\ha~ratios of Totoro. These comparisons suggest that shocks are unlikely to be the dominant mechanism that is responsible for ionizing the gas blob.

There have been studies showing that AGN is able to ionize gas clouds extending to several kpc \citep{lin09,fu08}, and the effect from AGN  may persist even $\sim10^{5}$yr after the central engines shut off \citep[e.g., the `Hanny's Voorewerp';][]{lin09}). If Totoro is indeed a `dark' galaxy or a cloud interacting with Satsuki, the tidal field would cause a gas inflow that fuels the central black hole of Satsuki and possibly trigger the AGN activity. 
Although no X-ray point source is detected in the position of either Satsuki or Totoro, the detection of point-like radio source (see Sec. 2.3) in the center of Satsuki indicates the presence of a low-activity AGN. Therefore, an alternative explanation for the line ratios seen in Totoro is due to the star formation -- AGN mixing effect \citep{dav14a,dav14b}. 

Observationally a starburst--AGN mixing sequence is often found in starbursting galaxy that hosts a central AGN, in which case the line ratio moves from Seyfert to composite to HII regions as the distance from the central AGN increases. The position of line ratios depends on the fractional contribution between AGN and star formation. According to the mixing model by \citet{dav14b}, when the star-forming cloud is  photonionized by an AGN, the line ratios can fall into the `composite' region on BPT diagrams. In the case of 100\% contribution from AGN, the emission line ratio falls into the `AGN' region, instead of `composite' region where Totoro lies. This implies that the emission line ratios seen in Totoro can not be fully attributed to pure AGN photonionization, and some level of star formation may be required. Unlike the typical star formation -- AGN mixing \citep{dav14a,dav14b}, our case is analogous to the star formation -- LINER sequence, in which the LINER excitation is due to the low-activity AGN located in Satsuki.

\subsection{Comparison to similar objects in the literature}
There are several similar systems reported in the literature that show offset ionized gas components, and hence it is intriguing to compare this Totoro with previous cases. For example, the famous Hanny's Voorewerp \citep{lin09}, also exhibits a separate warm gas component that is offset from the main galaxy by several kpc.  However, there are several different features between Hanny's Voorewerp and Totoro presented in this work. First, the ionized gas of Hanny's Voorewerp has much higher excitation state and its line ratios are consistent with Seyfert, as opposed to the `composite' region for Totoro. Moreover, the \ha~morphology of the Hanny's Voorewerp is lumpy and irregular, unlike a disk-like structure seen in our case. Another well-known example is an \ha~emission line component (often referred as the `cap') at a projected distance of 11 kpc northwest of M82 \citep{dev99,leh99,ste03}. The cap has a shell-like structure and may possibly be bow shock formed by the starburst-driven superwind. In both cases, the nearby galaxy is a late-type star forming galaxy, different from Satsuki
, which is an early-type galaxy. Totoro may thus represent a different category of offset ionized gas in nearby galaxies. 

\section{CONCLUSION}\label{sec:conclusion}

Here we report a discovery of a puzzling giant \ha~blob, Totoro, identified from the first-year MaNGA data. The data disfavor the scenario that Totoro is tidally-stripped  gas from MaNGA 1-24145 (Satsuki) that is interacting with the southern companion (Mei), or the ram-pressure stripped gas when Satsuki falls into the center of the cluster it is located. Despite there being no X-ray point source or radio jet detected in this system, we can not rule out the possibility that Totoro is ejected from a past AGN activity in Satsuki, which likely hosts a faint AGN given its radio luminosity. On the other hand, the \ha~morphology and the lack of stellar tidal streams suggest that Totoro could also be a separate `dark' galaxy (or an extremely LSB galaxy) interacting with Satsuki. The non-detection of the stellar continuum suggests Totoro is different from known dwarf populations or UDGs: it is either completely `dark' or with a star formation rate that contributes $<37\%$ of the \ha~flux. 

As for the source that powers the line excitation for Totoro, the `composite' line excitation can be explained either by a star-forming cloud being excited by a low-velocity shock or by the star formation -- LINER mixing effect. The shock scenario, however, is less favoured because of the low velocity dispersion observed in Totoro region. The decrease in the \othree/\hb~ratio away from Satsuki indicates that the ionizing source is possibly located inside Satsuki. Thus, the star formation -- LINER mixing effect seems to be the most probable ionizing mechanism. In this scenario, the hypothesis is that the ionizing source is the low-activity AGN residing in Satsuki, being triggered by the gas inflow induced during the interaction between Satsuki and the `dark' galaxy (or gas cloud). The AGN subsequently photoionizes the `dark' gas cloud, which then emits the \ha~photons. 

However, we have not considered the case where Totoro is tidally-stripped from Satsuki while falling into the cluster center at the same time, which results in the non-typical tidally or ram-pressure stripped gas morphology of Totoro. More sophisticated modelling considering both the effects of tidal disrupting and the orbital motions of Satsuki relative to the cluster environments, as well as future resolved atomic and molecular gas observations, such as HI and CO, are required to further understand the origin of Totoro.

\acknowledgments

We thank the anonymous referee for constructive suggestions which significantly improve the clarity of this paper. L. Lin thank I-Ting Ho, Michal Michalowski, Yen-Ting Lin, You-Hua Chu, Lisa Kewley, Tomo Goto, Jorge Barrera-Ballesteros, and Christy Tremonti for useful discussions. The work is supported by the Ministry of Science \& Technology of Taiwan
under the grant MOST 103-2112-M-001-031-MY3. H.F. acknowledges support from the NSF grant AST-1614326 and funds from the University of Iowa. S. Peirani acknowledges support from the Japan Society for the Promotion of
Science (JSPS long-term invitation fellowship). J.G.F-T is currently supported by Centre National d'Etudes Spatiales (CNES) through PhD grant 0101973 and the R\'egion de Franche-Comt\'e and by the French Programme National de Cosmologie et Galaxies (PNCG). 

Funding for the Sloan Digital Sky Survey IV has been
provided by the Alfred P. Sloan Foundation, the U.S.
Department of Energy Office of Science, and the Participating Institutions. SDSS-IV acknowledges support
and resources from the Center for High-Performance
Computing at the University of Utah. The SDSS web
site is www.sdss.org. SDSS-IV is managed by the Astrophysical Research Consortium for the Participating
Institutions of the SDSS Collaboration including the
Brazilian Participation Group, the Carnegie Institution
for Science, Carnegie Mellon University, the Chilean
Participation Group, the French Participation Group,
Harvard-Smithsonian Center for Astrophysics, Instituto
de Astrof\'isica de Canarias, The Johns Hopkins University, Kavli Institute for the Physics and Mathematics of the Universe (IPMU) / University of Tokyo, Lawrence
Berkeley National Laboratory, Leibniz Institut f\"ur Astrophysik Potsdam (AIP), Max-Planck-Institut f\"ur Astronomie (MPIA Heidelberg), Max-Planck-Institut f\"ur
Astrophysik (MPA Garching), Max-Planck-Institut f\"ur
Extraterrestrische Physik (MPE), National Astronomical Observatory of China, New Mexico State University,
New York University, University of Notre Dame, Observat\'ario Nacional / MCTI, The Ohio State University,
Pennsylvania State University, Shanghai Astronomical
Observatory, United Kingdom Participation Group, Universidad Nacional Aut\'onoma de M\'exico, University of
Arizona, University of Colorado Boulder, University of
Oxford, University of Portsmouth, University of Utah,
University of Virginia, University of Washington, University of Wisconsin, Vanderbilt University, and Yale University. 

The National Radio Astronomy Observatory is a facility of the National Science Foundation operated under cooperative agreement by Associated Universities, Inc.  This work used data from project AGBT16A\_095: `HI-MaNGA: HI Followup of MaNGA galaxies, PI Karen L. Masters. 

This work is partly based on observations obtained with MegaPrime/MegaCam, a joint project of CFHT and CEA/DAPNIA, at the Canada-France-Hawaii Telescope (CFHT) which is operated by the National Research Council (NRC) of Canada, the Institut National des Science de l\'Univers of the Centre National de la Recherche Scientifique (CNRS) of France, and the University of Hawaii.

This research has made use of data obtained from the Chandra Data Archive and the Chandra Source Catalog, and software provided by the Chandra X-ray Center (CXC) in the application packages CIAO, ChIPS, and Sherpa.

%\appendix

\end{document}